\documentclass[11pt, oneside]{article}

\usepackage[paper=a4paper, total={16cm,23cm}]{geometry}
\usepackage[T1]{fontenc}
\usepackage{setspace} % line-spacing
\usepackage[hang]{footmisc} % footnotes
\usepackage{titling} % title page
\usepackage{tocloft} % table of contents
\usepackage{parskip} % paragraph indentation

\usepackage{amsfonts}
\usepackage{amsmath}
\usepackage{amssymb}
\usepackage{mathrsfs, dsfont, amsthm}
\usepackage{physics}
\usepackage{mathtools}
\numberwithin{equation}{section} % numbering equations

\usepackage{xcolor}
\definecolor{dark-red}{rgb}{0.50,0.12,0.12} % links
\definecolor{mblue}{rgb}{0.30, 0.45, 0.70}
\definecolor{mred}{rgb}{0.70, 0.20, 0.20}
\definecolor{mgray}{rgb}{0.63, 0.63, 0.63}

\usepackage{hyperref} % hyperlink refs
\hypersetup{colorlinks=true, allcolors=dark-red, linktoc=all}

\usepackage{cite}

\usepackage{tikz}
\usepackage{tikzlings}
\usetikzlibrary{calc, patterns, decorations, decorations.markings, shapes, positioning, intersections, quotes, angles}
\usepackage[subrefformat=parens]{subcaption}
\usepackage{pgfplots}
\pgfplotsset{compat=newest}
\newcommand{\mathdefault}[1][]{}

\def \cO {\mathcal{O}}
\def \rU {\mathrm{U}}
\def \rmh {\mathrm{h}}
\def \bbR {\mathbb{R}}
\def \bbT {\mathbb{T}}
\def \bbZ {\mathbb{Z}}

\newcommand{\ep}{\mathrm{e}}
\newcommand{\ic}{\mathrm{i}}
\newcommand{\diff}{\mathrm{d}}
\newcommand{\AdS}{\text{AdS}}
\newcommand{\rh}{r_\mathrm{h}}
\newcommand{\rel}{\mathrm{rel}}
\newcommand{\mx}{\mathrm{max}}
\newcommand{\Arg}[1]{\text{Arg}\,#1}

\newcommand{\overbar}[1]{\mkern 1.3mu\overline{\mkern-1.3mu#1\mkern-1.3mu}\mkern 1.3mu}
\newcommand{\conj}[1]{\mkern 1.3mu\overline{\mkern-1.3mu#1\mkern-1.3mu}\mkern 1.3mu}

\begin{document}
% -------------------------
% title page
% -------------------------
\begin{titlingpage}
    \vspace*{3em}
    \onehalfspacing
    \begin{center}
        {\LARGE A new observable for holographic cosmology}
    \end{center}
    \singlespacing
    \vspace*{2em}
    \begin{center}
        % authors
        \textbf{
        Joydeep Chakravarty,$^1$
        Alexander Maloney,$^1$
        Keivan Namjou,$^1$
        and Simon F. Ross$^2$
        }
    \end{center}
    \vspace*{1em}
    \begin{center}
        \textsl{
        % affiliations
        $^1$\ Department of Physics, McGill University \\
        Montr\'eal, QC H3A 2T8, Canada \\[\baselineskip]
        }
        \textsl{
        % affiliations
        $^2$\ Centre for Particle Theory, Department of Mathematical Sciences, Durham University \\
        South Road, Durham DH1 3LE, UK \\[\baselineskip]
        }
        % emails
        \href{mailto:joydeep.chakravarty@mail.mcgill.ca}{\small joydeep.chakravarty@mail.mcgill.ca},
        \href{mailto:alexander.maloney@mcgill.ca}{\small alex.maloney@mcgill.ca} \\
        \href{mailto:keivan.namjou@mail.mcgill.ca}{\small keivan.namjou@mail.mcgill.ca},
        \href{mailto:joydeep.chakravarty@mail.mcgill.ca}{\small s.f.ross@durham.ac.uk}
    \end{center}
    \vspace*{3em}
    \begin{abstract}
    The double-cone geometry is a saddle of the gravitational path integral, which explains the chaotic statistics of the spectrum of black hole microstates. This geometry is the usual AdS-Schwarzschild black hole, but with a periodic identification of the time coordinate; the resulting singularity at the black hole horizon is regulated by making the geometry slightly complex. Here, we consider generalizations of the double-cone geometry which include the Lorentzian cosmology that sits between the event horizon and the black hole singularity. We analyze this in two and three dimensions, where the cosmology has compact spatial sections and big bang/crunch singularities. These singularities are regulated in the same way by slightly complexifying the metric. We show that this is possible while satisfying the Kontsevich-Segal criterion, implying that these geometries can be interpreted as perturbatively stable saddle points in general relativity. This procedure leads to a novel description of the cosmology in terms of standard observables in the dual boundary CFT. In three dimensions, the cosmological solution gives a new contribution to the two-point function of the density of states in the boundary CFT. Unlike the usual double cone, it describes correlations between black hole microstates with different masses, and in a limit describes correlations between the statistics of heavy states and states near the BTZ threshold.
    \end{abstract}
\end{titlingpage}
\tableofcontents

% -------------------------
% introduction
% -------------------------
\section{Introduction}
The study of quantum cosmology is one of the most difficult---and important---problems in theoretical physics. One possible approach is to use the AdS/CFT correspondence, but one immediate problem is that AdS is not a cosmology (at least in the usual sense). Another possibility is to take a path integral approach and integrate over an appropriate space of metrics. This is in principle elegant, but in practice difficult to implement and often plagued with ambiguities. One important exception is in low-dimensional systems, where in non-cosmological contexts controlled computations are possible and can be tested against boundary CFT expectations (most notably in two dimensions \cite{Saad:2019lba}). In general, however, we are restricted to the study of the saddle points' contributions to the path integral, i.e., solutions to Einstein's equations. Studies of Euclidean saddle points have led to important insights, for example, on black hole thermodynamics, but real Euclidean saddle points exist only in certain special cases; in more general settings, the saddle points that appear are complex.  The interpretation of such complex saddle points is often subtle, but here again, AdS/CFT provides valuable guidance.

In constructing our theory of cosmology, our inspiration is the double-cone geometry of \cite{Saad:2018bqo}. In the context of two-dimensional (Jackiw-Teitelboim) gravity, the metric of the double cone is that of AdS$_2$ metric in Rindler (i.e., black hole) coordinates:
\begin{equation}\label{eq:ads2r}
    \diff s^2 = - \sinh^2 u \, \diff t^2 + \diff u^2,
\end{equation}
with the time coordinate $t$ identified periodically, $t \sim t + T$.   This geometry has two boundaries at $u\to \pm\infty$, each of which is a time-like circle with period $T$.  So this geometry is interpreted as giving a contribution to the spectral form factor\footnote{The difference in signs comes from the fact that the isometry $\partial_t$ is future-directed in one asymptotic region and past-directed in the other; the saddle is topologically a cylinder, and a rotation around the cylinder will appear clockwise viewed from one end and anticlockwise from the other.} 
\begin{equation}\label{eq:sff}
    \overbar{Z(\beta + \ic T) Z(\beta - \ic T)},
\end{equation} where $Z(\beta+\ic T) = \Tr \ep^{-\beta H + \ic H T}$ is the partition function of the theory. More precisely, it  contributes to the square of a microcanonical version of the partition function,
\begin{equation}\label{eq:smear}
    Y_{E,\Delta E}(T) = \int \diff \beta \, \ep^{\beta E + \beta^2 \Delta E^2} Z(\beta + \ic T).
\end{equation}
Crucially, the overline in equation \eqref{eq:sff} indicates that this should be interpreted as an \textit{averaged} observable, a notion that can be understood precisely in theories with holographic duals.  In holographic theories, the averaging is a disorder average over a suitable family of microscopic boundary Hamiltonians (as was made completely precise in \cite{Saad:2019lba}).  In general, however, it can be interpreted as a more familiar coarse-graining over a small energy window as in equation (\ref{eq:smear}).

The double-cone geometry has a singularity at the tip of the cone, $u = 0$. So, one must proceed carefully in determining whether it is a legitimate contribution to the path integral.  One important point of evidence in favor of its inclusion is that, as noted in \cite{Saad:2018bqo}, its action can be computed and has an important physical interpretation.  Specifically, the classical action vanishes, and the one-loop determinant is linear in $T$, due to the mode which describes a (relative) reparametrization of the two boundary circles.  This linear growth in the spectral form factor is a generic expectation of a chaotic theory, where it describes the repulsion of energy eigenvalues.\footnote{Writing $Z(\beta)=\int \diff E \, \rho(E) \ep^{-\beta E}$ a linear growth in the form factor implies the typical repulsion $\overline{\rho(E) \rho(E')} \sim \frac{1}{(E-E')^2}$ expected for chaotic systems in the GUE universality class.}  This linear growth was dubbed the ``ramp'' in \cite{Cotler:2016fpe}, and appears generically in chaotic systems at late time. It is remarkable that this detailed spectral structure is encoded in a semi-classical gravitational saddle point.

For this to work, one needs to better understand the singularity at $u=0$. One approach is to resolve the singularity by complexifying the metric slightly; we will review this complexification (and others) in the next section. When one does so, the two asymptotic regions are connected by a smooth (although complex) geometry near $u=0$. The important point is that there is a clear criterion, known as the Kontsevich-Segal (KS) condition, that characterizes the precise circumstances under which a complex geometry can be viewed as a perturbatively stable saddle point \cite{Kontsevich:2021dmb, Witten:2021nzp}. This complexification procedure allows one to understand exactly which perturbative fluctuations are allowed in the double cone geometry \cite{Chen:2023hra}.\footnote{Indeed, the role of complex saddle points has been recognized elsewhere, including rotating black holes and cosmology \cite{Hartle:1983ai, Halliwell:1988ik, Halliwell:1989dy}, global symmetries in two dimensions \cite{Bah:2022uyz}, JT gravity in nearly dS \cite{Maldacena:2019cbz}, and holographic Schwinger-Keldysh contours \cite{Glorioso:2018mmw, Chakrabarty:2019aeu, Jana:2020vyx}. Most recently, this includes constraints on inflation \cite{Hertog:2023vot, Maldacena:2024uhs} and more general aspects of cosmology \cite{Lehners:2021mah, Jonas:2022uqb}.}

Our goal is to explore other complexifications of the double-cone geometry, to see what classical solutions they describe, and to investigate their interpretation as saddle points for observables in the dual CFT.  An important motivation for this exploration is the observation that inside the black hole horizon, a periodic identification of the time direction produces a spatially compact cosmology. This can be seen directly from equation (\ref{eq:ads2r}), where this interior region corresponds to imaginary values of $u$ so that $t$ is now a (compact) spatial direction. Such geometries have big bang/crunch singularities, which we will resolve by complexifying the geometry just as in the case of the double cone.  Remarkably, this complexification will connect the cosmology to the region outside the horizon---roughly, the big bang/crunch singularities will be replaced by asymptotically AdS regions with periodic Lorentzian time.  This offers a novel perspective on the quantum theory of the Lorentzian cosmology. This is particularly interesting in the context of three-dimensional gravity, where the resulting cosmological spacetime inside the horizon of BTZ is a torus cosmology, which has been extensively investigated.\footnote{ See \cite{Carlip:1998uc} for a review and \cite{Godet:2024ich, Banihashemi:2024aal} for more recent works.}

There are a number of antecedents for this construction: the Hartle-Hawking no-boundary proposal constructs cosmological saddle points by connecting a smooth compact Euclidean section to a Lorentzian cosmology \cite{Hartle:1983ai}. A similar recent construction with a negative cosmological constant connects a Euclidean AdS space with a boundary to a Lorentzian cosmology \cite{VanRaamsdonk:2020tlr}. Our construction is different in that these constructions connect a Euclidean solution to a Lorentzian one at a moment of time symmetry in the cosmology, while our construction will connect two Lorentzian solutions close to the singularity in the cosmology. As in the previous constructions, these saddle points could be used to define a wavefunction on a spatial slice of the cosmology. Our focus, however, is on understanding their description from the dual CFT perspective.

This CFT interpretation is possible because the solutions have two Lorentzian asymptotically AdS boundaries, as in the double cone. So, they can be interpreted in terms of the statistics of the density of states in the dual CFT. We will consider this in AdS$_2$ and AdS$_3$. In the AdS$_2$ case, we can construct contours that satisfy the KS condition, but they will not satisfy the same boundary conditions for the dilaton at both ends, so they are not contributions to a spectral form factor in a single theory.\footnote{We thank Martin Sasieta for pointing out the importance of the dilaton here.} In AdS$_3$, the new contours give contributions to a novel observable that describes correlations between the spectrum at \textit{different} values of the energy in a single theory. Our main result is that the torus cosmology contributes to the spectral observable
\begin{equation}\label{eq:microavg}
    \overbar{Y_{\tilde{E},\Delta \tilde{E}}(\tilde{T}) Y_{E, \Delta E}(-T)},
\end{equation}
where $\tilde{E} = \frac{T^2}{4\pi^2} E$ and $\tilde{T} = \frac{4\pi^2}{T}$. Here we are normalizing our energy so that empty AdS has negative energy $E=-\frac{\ell}{8 G}$, and $E\to 0$ corresponds to states just at the threshold for black hole formation in AdS$_3$. As with the double cone, this observable is sensitive to pair correlations between the energy eigenvalues. The difference, however, is that we will now make a prediction for the correlations between eigenvalues at \textit{different} values of the energy. If one takes large $T$ with fixed $E$, this describes correlations between states with very high energy (i.e., black hole microstates) and states in the bulk of the black hole spectrum.  On the other hand, if we take large $T$ with fixed $\tilde{E}$ it describes correlations between black hole states and states with energy near 0. This is a surprising result from the usual random matrix theory perspective.

In the next section, we will describe in detail our geometries, and their complexification, and analyze the Kontsevich-Segal condition which guarantees their stability. This will lead us to the cosmological interpretation of our new spectral observable (\ref{eq:microavg}).  In section \ref{s:action} we will discuss the action and one-loop determinants on these solutions and make a specific prediction for this observable. In section \ref{s:fields} we analyze field theory in this background, which is described in terms of the spectrum of black hole (anti-)quasinormal modes.  We conclude in section \ref{s:discussion} with some speculation and discussion of future directions. Appendix \ref{s:kshd} contains comments on the higher-dimensional cases.

% -------------------------
% contours
% -------------------------
\section{Cosmology from complex geometries}\label{s:contours}
In this section, we describe the complex geometries that probe both the exterior and the interior of the AdS-Schwarzschild geometry. This family of geometries includes the double cone of \cite{Saad:2018bqo} (which describes the ``ramp'' in the spectral form factor) as well as novel geometries that probe the cosmological region behind the event horizon.

We begin by considering, in general, an AdS-Schwarzschild black hole in $d+1$ dimensions, with the following metric:
\begin{equation}
    \diff s^2 = - f(r) \, \diff t^2 + \frac{\diff r^2}{f(r)} + r^2 \, \diff \Omega_{d-1}^2,
\end{equation}
where $f(r) = r^2 + 1 - \frac{M}{r^{d-2}}$, using units where the AdS radius $\ell_\AdS = 1$. This has a horizon at $r = \rh$, where $f(\rh) = 0$. The double-cone geometry is found by identifying $t \sim t + T$ in the two exterior regions with $r > \rh$. In the interior, $r < \rh$, $f(r) < 0$, and this metric describes a time-dependent spacetime. After we identify $t \sim t + T$ this is a cosmology with compact spatial sections. Before making the identification, the exterior and interior regions are smoothly connected, but the identification in $t$ has a fixed point at the bifurcation surface, so the relation is a little more complicated. Nonetheless, there is clearly a close relation between the cone geometry outside the horizon and the cosmology, which we will exploit to understand better the quantum theory in cosmological contexts.

To do so, we will complexify the metric, which is most clearly described in an isotropic coordinate
\begin{equation}
   u = \int_{\rh}^r \frac{\diff r'}{\sqrt{f(r')}}.
\end{equation}
where the metric takes the general form
\begin{equation}
    \diff s^2 = - g(u) \, \diff t^2 + \diff u^2 + h(u) \, \diff \Omega_{d-1}^2.
\end{equation}
This coordinate is sometimes introduced as a convenient way to describe the whole Einstein-Rosen bridge, as $u \in (-\infty, \infty)$ covers the two exterior regions. It thus provides a good coordinate system for the double-cone geometry. In the cosmology, since $f(r) < 0$, $u$ becomes imaginary, and the cosmology inside the horizon corresponds to (a portion of) the imaginary axis in the complex $u$-plane.

As reviewed in the introduction, even for the double-cone geometry, one must complexify the metric in order to avoid the singularity at $u = 0$, where $g(0) = 0$. This complexification can be taken to be arbitrarily small and thus should really be viewed as an infinitesimal regularization of the geometry which allows one to properly treat the double cone as a saddle point of general relativity. The appearance of the cosmology on the imaginary axis further motivates us to investigate other possible contours in the complex $u$-plane. Indeed, we will discover that it is similarly possible to make an arbitrarily small complex deformation of the cosmology that is obtained when $u$ is purely imaginary. In a similar way, this regularizes the cosmological ``big bang'' and ``big crunch'' singularities. Once we do so, we will discover something remarkable: an interpretation of this cosmology in terms of a particular CFT observable (a spectral two-point function) in the boundary theory.

The complex geometries we will consider are obtained by specifying a contour in the complex $u$-plane, which we will denote $u(\rho)$ with $\rho$ real. The metric on this complex geometry is
\begin{equation}
    \diff s^2 = -g(u) \, \diff t^2 + \qty(\frac{\partial u}{\partial \rho})^2 \, \diff \rho^2 + h(u) \, \diff \Omega_{d-1}^2.
\end{equation}
Because this is related by a (complex) change of coordinates to the original AdS-Schwarzschild geometry, this complex geometry will continue to be a solution of Einstein's equation for any choice of contour $u(\rho)$. The original double-cone geometries correspond to the contour $u=\rho+\ic \epsilon$ for some small $\epsilon$.
Our goal is to investigate contours that look like cosmologies.

Of course, not every complex metric will be allowed. We will demand that our metrics obey the Kontsevich-Segal criterion \cite{Kontsevich:2021dmb, Witten:2021nzp}, which requires that the real part of the matrix $\sqrt{\det g} g^{\mu\nu}$ be positive definite. For a diagonal metric $g_{\mu\nu} = \lambda_\mu \delta_{\mu\nu}$, this is the same as the statement that the sum of the complex phases of the metric components is less than $\pi$:
\begin{equation}\label{eq:ksc}
    \sum_{\mu=1}^{d+1} \abs{\Arg{\lambda_\mu}} < \pi.
\end{equation}
Metrics that satisfy the Kontsevich-Segal criterion have the property that perturbative field theories based on form fields are well defined, in the sense that all of the Gaussian integrals that appear when one expands a field theory around this background are convergent. So it appears to be a necessary (though possibly not sufficient) criterion for a metric to be treated as a well-defined and perturbatively stable saddle-point contribution to a gravitational path integral.

This will constrain significantly the possible choices of contour, in a way that depends non-trivially on dimension. In the remainder of this section, we will study the allowed contours in two and three dimensions; we relegate a discussion of higher dimensions to an appendix. In section \ref{s:ks2d}, we study two dimensions, where the contours include a Hartle-Hawking-like contour which interpolates between a Euclidean and a Lorentzian geometry, and new contours that contribute to the spectral form factor, interpolating between Lorentzian geometries. In section \ref{s:ks3d}, we study three dimensions, where we find contours that satisfy the Kontsevich-Segal criterion and interpolate between two Lorentzian cones whose cycles are swapped compared to each other. These are the main results of our paper.

\subsection{Warm-up: new contours in two dimensions}\label{s:ks2d}
In two dimensions, we write AdS$_2$ in the isotropic black hole coordinates as
\begin{equation}\label{eq:dcme}
    \diff s^2 = - \sinh^2 u \, \diff t^2 + \diff u^2.
\end{equation}
The double-cone geometry of \cite{Saad:2018bqo, Chen:2023hra} is obtained by identifying $t \sim t + T$ and considering real $u \in (-\infty,\infty)$. Since this geometry is real and Lorentzian, it saturates the Kontsevich-Segal bound. We can then regulate the geometry by taking $u=\rho + \ic \epsilon$. The resulting geometry then obeys the Kontsevich-Segal condition \cite{Witten:2021nzp}.

In the complex $u$-plane, we have other interesting solutions: if we set $u = \rho \pm \ic \frac{\pi}{2}$, we find a Euclidean AdS$_2$ metric. As a real Euclidean metric, this of course also obeys the Kontsevich-Segal condition. This was exploited in \cite{Chen:2023hra} to understand the modified time-translation generator associated with the version of the double cone, where we go around the singularity at $\rho=0$ in the complex plane. Similarly, if we set $u = \rho \pm \ic \pi$, we obtain the Lorentzian AdS$_2$ space-time again.

Our primary interest, however, is the solution where $u$ is purely imaginary: $u = \ic \rho$. This describes a cosmology that runs from a big bang at $\rho = 0$, reaching a maximum size where it crosses the Euclidean section at $\rho = \pm \frac{\pi}{2}$, and ending with a big crunch at $\rho = \pm \pi$. As this geometry saturates the Kontsevich-Segal criterion, we would like to understand whether it can be regulated in the same way as the double-cone geometry.

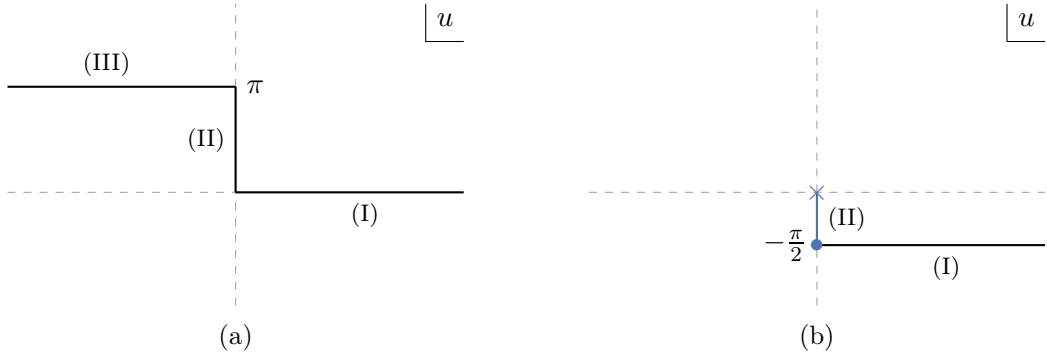
\begin{figure}
\centering
\begin{subfigure}{.47\textwidth}
    \centering
    \begin{tikzpicture}[scale=1]
        \draw[dashed,mgray] (-3,0)--(0,0) (0,-1.5)--(0,0) (0,1.4)--(0,2.5);
        \draw (2.5,2.5)--(2.5,2)--(3,2) (3,2.5)node[anchor=north east]{$u$};
        \draw[thick] (0,0)--(1.7,0)node[anchor=north]{\footnotesize (I)}--(3,0);
        \draw[thick] (0,0)--(0,.7)node[anchor=east]{\footnotesize (II)}--(0,1.4)node[anchor=west]{$\pi$};
        \draw[thick] (0,1.4)--(-1.7,1.4)node[anchor=south]{\footnotesize (III)}--(-3,1.4);
    \end{tikzpicture}
    \caption{}\label{fig:2d-uplane-dc}
\end{subfigure}
\begin{subfigure}{.47\textwidth}
    \centering
    \begin{tikzpicture}[scale=1]
        \draw[dashed,mgray] (-3,0)--(3,0) (0,-1.5)--(0,-.7) (0,0)--(0,2.5);
        \draw (2.5,2.5)--(2.5,2)--(3,2) (3,2.5)node[anchor=north east]{$u$};
        \draw[thick] (0,-.7)--(1.7,-.7)node[anchor=north]{\footnotesize (I)}--(3,-.7);
        \draw[thick] (0,-.7)node[anchor=east]{$-\frac{\pi}{2}$} (0,-.35)node[anchor=west]{\footnotesize (II)};
        \draw[thick,mblue] (0,-.7)node{$\bullet$}--(0,0)node{$\times$};
    \end{tikzpicture}
    \caption{}\label{fig:2d-uplane-hh}
\end{subfigure}
    \caption{Contours in the $u$-plane for AdS$_2$: contour (a) with the same metric asymptotics as the double cone, and contour (b), where $\Im u = \pm \frac{\pi}{2}$ in region (I), which leads to the Hartle-Hawking state (${\color{mblue} \bullet}$) at $u = \pm \ic \frac{\pi}{2}$, ending up at a big crunch (${\color{mblue} \times}$) at $u=0$ after evolving through the cosmology in region (II).}\label{fig:2d-uplane-gen}
\end{figure}

\begin{figure}
\centering
    \input{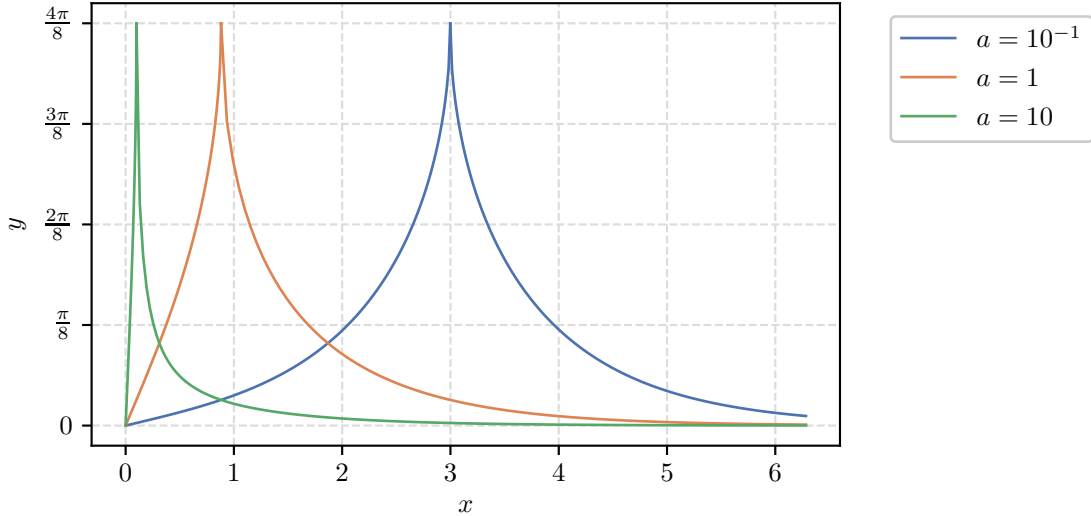}
    \caption{Critical contours which saturate the Kontsevich-Segal condition, plotted in the complex $u$-plane for the AdS$_2$ case.}\label{fig:2d-contours}
\end{figure}

Let us consider a general contour $u(\rho)$, and write $u=x + \ic y$. The Kontsevich-Segal criterion (\ref{eq:ksc}) becomes
\begin{equation}\label{eq:kscv-2d}
    2 \arctan \abs{\frac{\diff y}{\diff x}} + 2 \arctan \qty(\tan(\frac{\pi}{2} - y)  \tanh \abs{x}) < \pi,
\end{equation}
which restricts the allowed angles of the contour in the complex $u$-plane at each point. To understand the range of possibilities, it is useful to plot the \textit{critical contours}, those which have $\sum_{\mu=1}^2 \abs{\Arg{\lambda_\mu}} = \pi$. At each point, there are two such directions, distinguished by the sign of $\frac{\diff y}{\diff x}$. These satisfy
\begin{equation}
    \frac{\diff y}{\diff x} = \pm \tan(\frac{\pi}{2} - \arctan \qty(\tan (\frac{\pi}{2} - y) \tanh \abs{x})) = \pm \frac{\tan y}{\tanh x},
\end{equation}
with solutions (corresponding to the plus and minus signs respectively)
\begin{equation}
    \sin y = a \sinh x, \qquad \text{and} \qquad \sin y = \frac{1}{a \sinh x}.
\end{equation}
These are plotted for $y \in (0, \frac{\pi}{2})$ in figure \ref{fig:2d-contours}. The contours are symmetric about $y=\frac{\pi}{2}$ so the contours for $y \in (\frac{\pi}{2}, \pi)$ are obtained by reflection. As expected, we see that the direction of the contour is unrestricted around the Euclidean AdS$_2$ at $y = \frac{\pi}{2}$. The critical contours that start at the origin return to $(x,y) = (0, \pi)$, while the ones that come in from infinity go back out to infinity along $y = 0$ and $y = \pi$. We see that the latter have $y \to 0$ as $x \to \infty$ for all values of $a$. The critical contours correspond to the case where the KS criterion is saturated. Hence, a physical contour's slope at each point should be bounded above and below by two critical contours to satisfy the KS criterion.

A familiar example of a contour involving the cosmology is a Hartle-Hawking-like contour shown in figure \ref{fig:2d-uplane-hh}, where we run along the Euclidean solution at $y = \pm \frac{\pi}{2}$, and then along a portion of the imaginary axis, defining a state in the cosmology.

We are interested instead in contours that asymptotically approach the cone geometries. We note that there is a Lorentzian cone geometry at $y=0$, but there is also another one at $y=\pm \pi$, and we can consider a contour that runs between them. One option for such a contour is the one pictured in figure \ref{fig:2d-uplane-dc}, running through the cosmology. There is, however, a family of other possibilities consistent with the Kontsevich-Segal criterion.

From the boundary perspective, these contours have the same metric asymptotics as the double cone. However, if we consider these as solutions in JT gravity, they have different asymptotics for the dilaton; \eqref{eq:dcme} is a solution with $\phi = \phi_\rmh \cosh u$, so if the dilaton has a positive value at real $u$ it will have a negative value for $u = \rho + \ic \pi$. Thus, these contours {\it do not} contribute to the spectral form factor in JT gravity.\footnote{We thank Martin Sasieta for pointing out this issue with the dilaton.}

\subsection{Contours in three dimensions}\label{s:ks3d}
We now consider our main interest, complex contours in three dimensions. We start with the BTZ black hole in isotropic coordinates:
\begin{equation}\label{eq:btz-metric}
    \diff s^2 = - \rh^2 \sinh^2 u \, \diff t^2 + \diff u^2 + \rh^2 \cosh^2 u \, \diff \phi^2,
\end{equation}
where $\rh$ is the horizon radius. We identify $\phi \sim \phi+2 \pi$ and $t \sim t+T$, with the latter being the identification that gives the double cone. The boundary CFT lives on a Lorentzian torus.

As before, the usual double cone represents a contribution to
\begin{equation}\label{eq:3d-ucontour-0}
    \overbar{\abs{Y_{E, \Delta E}(T)}^2}, \qquad \text{with} \qquad E=\frac{\rh^2}{8G}.
\end{equation}
Following \cite{Saad:2018bqo, Chen:2023hra}, the singularity at $u=0$ is resolved by shifting the contour in the complex $u$-plane to $u = \rho \pm \ic \epsilon$ for real $\rho$. In three dimensions,  $y = \ic \frac{\pi}{2}$ is a second copy of the Lorentzian geometry, but with $t$ and $\phi$ interchanged. In other words, setting $u = \tilde{\rho} + \ic \frac{\pi}{2}$ gives
\begin{equation}
    \diff s^2 = \rh^2 \cosh^2 \tilde{\rho} \, \diff t^2 + \diff \tilde{\rho}^2 - \rh^2 \sinh^2 \tilde{\rho} \, \diff \phi^2.
\end{equation}
This has period $2\pi$ in the timelike direction and period $T$ in the spacelike direction. To relate to more standard CFT observables, it is useful to make a conformal transformation to restore the standard periodicity in the spacelike direction. This corresponds to a coordinate transformation
\begin{equation}
    \phi = \frac{T}{2\pi} \tilde{t}, \qquad t = \frac{T}{2\pi} \tilde{\phi}.
\end{equation}
We can rewrite the resulting solution by defining $\rh = \frac{2\pi}{T} \tilde{r}_\rmh$, so the metric is now
\begin{equation}
    \diff s^2 = - \tilde{r}_\rmh^2 \sinh^2 \tilde{\rho} \, \diff \tilde{t}^2 + \diff \tilde{\rho}^2 + \tilde{r}_\rmh^2 \cosh^2 \tilde{\rho} \, \diff \tilde{\phi}^2,
\end{equation}
with $\tilde{t} \sim \tilde{t}+\tilde{T}$ and $\tilde{\phi} \sim \tilde{\phi}+2 \pi$, where $\tilde{T} = \frac{4\pi^2}{T}$. This metric saturates the Kontsevich-Segal criterion but can be regulated as before by subtracting a small imaginary part from $u$. We conclude that the contour $u = \tilde{\rho} + \ic \frac{\pi}{2}$ contributes to
\begin{equation}\label{eq:3d-ucontour-ip2}
    \overbar{\abs{Y_{\tilde{E}, \Delta E}\qty(\frac{4\pi^2}{T})}^2}, \qquad \text{with} \qquad \tilde{E} = \frac{T^2 \rh^2}{32 \pi^2 G}.
\end{equation}

The segment of the imaginary axis between these two contours, $u = \ic \xi$ for $\xi \in \qty(0, \frac{\pi}{2})$, is the torus cosmology:
\begin{equation}
    \diff s^2 = - \diff \xi^2  + \rh^2 \sin^2 \xi \, \diff t^2 + \rh^2 \cos^2 \xi \, \diff \phi^2.
\end{equation}
Of course, this geometry exactly saturates the Kontsevich-Segal condition.

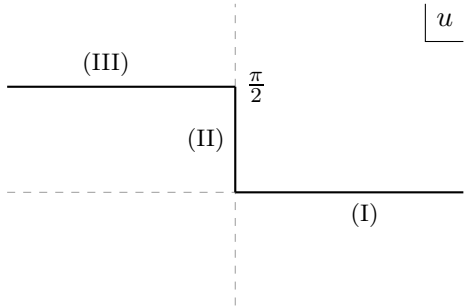
\begin{figure}
  \centering
    \begin{tikzpicture}[scale=1]
        \draw[dashed,mgray] (-3,0)--(0,0) (0,-1.5)--(0,0) (0,1.4)--(0,2.5);
        \draw (2.5,2.5)--(2.5,2)--(3,2) (3,2.5)node[anchor=north east]{$u$};
        \draw[thick] (0,0)--(1.7,0)node[anchor=north]{\footnotesize (I)}--(3,0);
        \draw[thick] (0,0)--(0,.7)node[anchor=east]{\footnotesize (II)}--(0,1.4)node[anchor=west]{$\frac{\pi}{2}$};
        \draw[thick] (0,1.4)--(-1.7,1.4)node[anchor=south]{\footnotesize (III)}--(-3,1.4);
    \end{tikzpicture}
    \caption{A contour in the $u$-plane for BTZ black hole: this contour passes through the cosmology in region (II), but connects on either end to half of the double-cone geometry in regions (I) and (III). Thus, the cosmological contour has two asymptotically AdS boundaries once we regulate the cosmological singularities by complexifying the metric.}\label{fig:3d-ucontour}
\end{figure}

Our goal is to deform this cosmological contour to satisfy the Kontsevich-Segal criterion. We construct a contour that approaches the horizontal contours asymptotically and passes through the cosmology, as sketched in figure \ref{fig:3d-ucontour}. This contour will offer some insight into the microscopic description of the cosmology.

To investigate the Kontsevich-Segal criterion, we again write $u = x + \ic y$ with $y \in \qty(0, \frac{\pi}{2})$, and the condition (\ref{eq:ksc}) becomes
\begin{equation}\label{eq:kscv-3d}
    2 \arctan \abs{\frac{\diff y}{\diff x}} + 2 \arctan \qty(\tan (\frac{\pi}{2} - y)  \tanh \abs{x}) + 2 \arctan \qty(\tan y \tanh \abs{x}) < \pi.
\end{equation}
The KS criterion restricts the angle of the contour in the complex $u$-plane at each point. In particular, for large $\abs{x}$, we find that the right-hand side of equation (\ref{eq:kscv-3d}) takes the form
\begin{equation}
    2 \arctan \qty(\tan (\frac{\pi}{2}- y) \tanh \abs{x}) + 2 \arctan (\tan y \tanh \abs{x}) = \pi - \cO\qty(\ep^{-2\abs{x}}).
\end{equation}
Consequently, all allowable contours are restricted to be nearly horizontal at large $\abs{x}$.

\begin{figure}
\centering
    \input{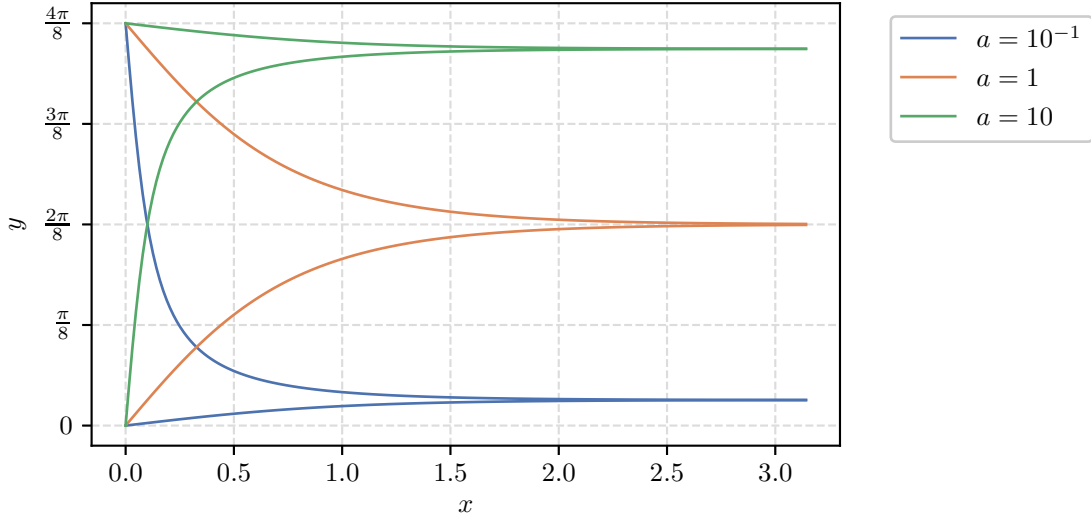}
    \caption{Critical contours in the complex $u$-plane in three dimensions.}\label{fig:3d-contours}
\end{figure}

As in the AdS$_2$ case it is useful to consider the critical contours with $\sum_{\mu=1}^3 \abs{\Arg{\lambda_\mu}} = \pi$, which delimit the range of allowed contours. At each point, there are two such directions, distinguished by the sign of $\frac{\diff y}{\diff x}$. These have
\begin{equation}
    \frac{\diff y}{\diff x} = \pm \tan(\frac{\pi}{2} - \arctan \qty(\tan(\frac{\pi}{2}- y) \tanh \abs{x}) -  \arctan \qty(\tan y \tanh \abs{x})) = \pm \frac{\sin 2y}{\sinh 2x},
\end{equation}
with solutions (corresponding to the plus and minus signs respectively)
\begin{equation}
    y = \arctan(a \tanh x), \qquad \text{and} \qquad y = \arctan (a \coth x),
\end{equation}
plotted in figure \ref{fig:3d-contours}. These critical contours are horizontal at large $x$, with the curves with $\frac{\diff y}{\diff x} >0$ approaching the origin as $x \to 0$, and the curves with $\frac{\diff y}{\diff x} <0$ approaching $(x,y) = \qty(0, \frac{\pi}{2})$. Similar to the previous case, the allowed contours lie in between the critical contours, with their slope at each point bounded above and below by the critical contours.

We wish to construct a contour that is within a distance $\epsilon$ from the real axis at large $x$. As $x$ becomes smaller, it cannot rise faster than the limiting contour with $\frac{\diff y}{\diff x} < 0$. A limit curve with $y \sim \epsilon$ asymptotically has $a \sim \epsilon$. This will remain close to the real axis until $x \sim \epsilon$. Thus, all allowed contours that start near the real axis hug the real axis until close to $x=0$. Similarly, a contour starting close to $y = \frac{\pi}{2}$ as $x \to -\infty$ stays horizontal until close to $x=0$. Thus, if we want a contour that is close to $y=0$ for $x \to \infty$ and close to $y = \frac{\pi}{2}$ for $x \to -\infty$, it will have to run between $0$ and $\frac{\pi}{2}$ close to the cosmology contour at $x \approx 0$.

There are many contours that would work. A simple explicit example is
\begin{equation}\label{eq:observm}
\begin{split}
    y  = \arctan \qty(\frac{\epsilon}{\epsilon +\tanh x}) \qquad \text{for} \qquad x>0, \\
    y = \frac{\pi}{2} - \arctan \qty(\frac{\epsilon}{\epsilon - \tanh x}) \qquad \text{for} \qquad x<0.
\end{split}
\end{equation}
Here we have demanded that the curve and its first derivative are smooth at $(x,y)=(0, \frac{\pi}{4})$, although the second derivative is not. However, the continuity of the first derivative is enough to make this a solution of Einstein's equations at $x=0$. One can further modify the contour near $x=0$ to smooth it out even more; there are no constraints on the contour near $x=0$ so it can be smoothed out in any desired way. We have thus constructed contours of the desired form that run through the cosmology and connect two asymptotic cone regions. Furthermore, we see that, unlike in the two-dimensional case, the space of possibilities with these asymptotics is quite limited, so understanding the dual observable offers us a better hope of probing the role of the cosmology.

On this contour, there are two asymptotic cone regions, one of which has period $T$ in the timelike direction and energy $E$, while the other has period $\tilde{T} = \frac{4\pi^2}{T}$ and energy $\tilde{E} = \frac{T^2}{4\pi^2} E$. From the boundary perspective, this would therefore represent a contribution to $$\overbar{Y_{\tilde{E},\Delta \tilde{E}}(\tilde{T}) Y_{E, \Delta E}(-T)},$$ as claimed in the introduction. This is a distinct observable, which is helpful as it cleanly separates the contribution of this saddle point from that of the usual double cone. It is also an interesting quantity from the dual perspective; a nonzero value for this observable indicates some pair correlation between energy eigenstates at different energies.

% -------------------------
% classical action
% -------------------------
\section{Gravity on complexified geometries}\label{s:action}
The contours described above represent novel saddle points in the path integral of general relativity. To understand their contribution to the spectral two-point function, we will now evaluate their classical action and consider the one-loop determinant around these saddles.

We will focus in this section on the more interesting case of three dimensions, where the action is the usual Einstein-Hilbert action, but a similar analysis would apply to the two-dimensional case, using the JT gravity action.

\subsection{Classical Einstein-Hilbert action}
For purely Euclidean or Lorentzian solutions, the calculation of the action is a standard exercise, but this is a little more non-trivial for our saddles, as they involve non-trivial contours in the complex plane. The prescription we will adopt for evaluating the action along such a contour is to evaluate the integrand (essentially the volume form, because our solution is locally AdS) on the real axis and continue it to the full contour as a holomorphic function of $u$. Note that because of this, any bulk contributions to the action will be independent of small deformations of the contour as long as one remains in a region where the integrand is analytic. Thus, the action we compute will turn out to be independent of the exact choice of contour through the bulk. 

To set the notation, let us first briefly review the action calculation in the three-dimensional Euclidean black hole, and in the Lorentzian region outside the horizon, corresponding to real $u$. In three dimensions with Euclidean signature, the gravity action is
\begin{equation}
    I = - \frac{1}{16\pi G} \int \diff^3x \, \sqrt{g} (R-2\Lambda) - \frac{1}{8\pi G} \int \diff^2x \, \sqrt{h} K + \frac{1}{8\pi G} \int \diff^2x \, \sqrt{h}.
\end{equation}
For simplicity, we will set $\ell_\AdS = 1$. The equations of motion give us $R-2\Lambda= 4 \Lambda = - \frac{4}{\ell^2} = -4$. For BTZ solution (\ref{eq:btz-metric}), $\sqrt{h} K = \rh^2 \cosh 2\rho$, and the volume is
\begin{equation}
    \int \diff^3x \, \sqrt{g} = 2\pi \beta \rh^2 \int_0^{\rho_\mx} \diff \rho \, \sinh \rho \cosh \rho = \frac{\pi}{2} \beta \rh^2 (\cosh 2 \rho_\mx - 1).
\end{equation}
In the action, the divergent part of the volume and the entire extrinsic curvature term cancel against the counterterm action, leaving just the finite term from the volume,\footnote{As a check, using $\beta = \frac{2\pi}{\rh}$, the action becomes $I = - \frac{(2\pi)^2}{8G \beta}$. This gives $S_{BH} = (\beta \frac{\partial}{\partial \beta} - 1) I =  \frac{(2\pi)^2}{4G \beta} = \frac{2\pi \rh}{4G}$ as expected. The energy is $E = \frac{\partial}{\partial \beta} I = \frac{\rh^2}{8G}$.}
\begin{equation}
    I = - \frac{\beta \rh^2}{8G}.
\end{equation}

In the Lorentzian action for the region outside of the black hole horizon, the volume form is $\sqrt{-g} = \rh^2 \sinh \rho \cosh \rho$ again, so the only difference is in the periodicity of $t$. The Lorentzian action for a single cone is thus
\begin{equation}\label{eq:singleconeaction}
    S =  -\frac{T \rh^2}{8G} = -ET.
\end{equation}

Consider now the standard double-cone contour, where we integrate along the real axis, passing underneath the singularity at $u=0$ in the complex $u$-plane. We define the integrand to be a holomorphic function of $u$, so the volume is
\begin{equation}
    \int \diff^3x \, \sqrt{-g} = 2\pi T \rh^2 \int_{-\rho_\mx}^{\rho_\mx} \diff \rho \, \sinh \rho \cosh \rho = 0,
\end{equation}
as the integrand is an odd function of $\rho$. Similarly, defining the volume measure and extrinsic curvature in the boundary term to be $\sqrt{-h} = \rh^2 \sinh \rho \cosh \rho$ and $\sqrt{-h} K = \rh^2 \cosh 2 \rho$ respectively, we see that two boundary terms will also cancel so that the overall value of the action vanishes.\footnote{It might not seem so clear that such an analytic continuation prescription is natural for the boundary terms, as these are only evaluated at the endpoints, not along the contour. However, the prescription for defining the boundary terms must be chosen consistently with the bulk prescription for the action to remain stationary under variations preserving the relevant boundary conditions.} This agrees with the claim in \cite{Saad:2018bqo} that the action should in general vanish for the double cone, as the two sides are complex conjugates of each other. Note that this is different from simply taking the action (\ref{eq:singleconeaction}) of the cone region outside the horizon twice, which is what we might be naturally led to do if we ignored the deformation of the contour near $\rho=0$.

In the contour introduced in the previous subsection, taking the integrand to be a holomorphic function of $u$, the action will also vanish. Traversing the contour in figure \ref{fig:3d-ucontour} along region (III) to region (I), the bulk contribution is given by:
\begin{equation}
\begin{split}
    S_{\mathrm{bulk}} &= \int_{-{\rho_\mx}}^0 \diff \rho_- \, \sinh (2 \rho_- + \ic \pi) - \ic \int_0^{\frac{\pi}{2}} \diff \xi \, \sinh(2 \ic \xi) + \int_0^{\rho_\mx} \diff \rho_+ \, \sinh 2 \rho_+ \\
&= - \frac{1}{2} \qty(1 - \cosh 2\rho_\mx) - \frac{1}{2} (\cos \pi -1) + \frac{1}{2} \qty(\cosh 2\rho_\mx -1) = \cosh 2\rho_\mx.
\end{split}
\end{equation}
In the boundary terms, we have $\sqrt{h} = \frac{1}{2} \sinh 2u$;  at $u = -\rho_\mx + \ic \frac{\pi}{2}$ we get $\sinh 2u = \sinh 2 \rho_\mx$, so the boundary terms at each end have the same sign in this contour, and the boundary contributions cancel the divergent part of the bulk, leaving us with zero action. In fact, this is in line with the intuition that the classical action would contribute as an overall phase of the form $\ep^{\ic \tilde{E} \tilde{T} - \ic E T}$ to the observable (\ref{eq:observm}): since $\tilde{E} \tilde{T} = ET$ in this geometry, the phase is zero.

Thus, the classical action for our three-dimensional contour vanishes, just as for the double-cone contour. Similarly in two dimensions, the JT gravity action will vanish along the contours interpolating between $y=0$ and $y=\pi$. The only significant difference in the two-dimensional case is that the bulk contribution to the action vanishes on-shell, so there are only boundary contributions.

\subsection{Gravitational one-loop determinant}
We are also interested in the one-loop determinant of the metric and bulk fields around this saddle point. An important contribution comes from the metric zero modes, as in \cite{Saad:2018bqo, Chen:2023hra}. There is a relative time shift $T_\rel$ between the two boundaries, which is canonically conjugate to the energy $E = \frac{\rh^2}{8G}$, i.e., $\qty{T_\rel, E} = 1$ \cite{Kuchar:1994zk, Harlow:2018tqv}. As we are in three dimensions, there is also a relative shift $\phi_\rel$ of the angular coordinate between the two boundaries, canonically conjugate to the angular momentum (we have set the angular momentum to zero in our discussion for simplicity). Thus, for fixed values of the energy and angular momentum, we actually have a two-parameter family of classical saddles to integrate over. The range of integration in $T_\rel$ is fixed by the periodicity $T$ we have chosen for the $t$ coordinate, and the range of integration of $\phi_\rel$ is similarly fixed by the $2\pi$ periodicity of $\phi$. More precisely, we also integrate over a window of energies and angular momenta, so the path integral includes a zero mode integral over
\begin{equation}
    \int_0^{\Delta E} \diff E \, \int_0^T \frac{\diff T_{\rel}}{2\pi} \qquad \text{and} \quad \int_0^{\Delta J} \diff J \, \int_0^{2\pi} \diff \phi_{\rel}.
\end{equation}
The first integral gives a contribution of the form $T \Delta E$, and the second integral gives a contribution of the form $2\pi \Delta J$. While we write the first integral as $T \Delta E$, the expression is actually symmetric between the two boundaries: while the periods $T$ and $\tilde{T}$ in the two asymptotic regions are not equal, the relation between the energies implies that $T \Delta E = \tilde{T} \Delta \tilde{E}$. Hence, integrating over this family of solutions gives us a factor of the volume of the torus in the one-loop determinant, $2\pi T$. As in the double cone \cite{Saad:2018bqo}, this implies that the contribution from these saddles grows linearly with $T$
\begin{equation}
    \overbar{Y_{\tilde{E},\Delta \tilde{E}}(\tilde{T}) Y_{E, \Delta E}(-T)} \simeq T
\end{equation}
This is a cosmological version of the ramp in the spectral form factor.  In the microscopic theory, we expect this ramp behavior to be cut off at some value of $T$ but this will not be visible in our semi-classical saddle-point analysis.

Because we chose a contour that satisfies the Kontsevich-Segal criterion, the remaining contribution to the one-loop determinant from nonzero modes will be well-behaved and finite. That is, we expect the spectrum of fluctuations around this contour to have negative imaginary parts so that their contribution to the one-loop determinant goes to one at large $T$. In the next section, we consider explicitly the fluctuations of a scalar field around our contours and find that the spectrum is related to the quasinormal modes of the black hole, as in \cite{Chen:2023hra}.

% -------------------------
% scalar field theory
% -------------------------
\section{Fields on complexified geometries}\label{s:fields}
To better understand our cosmological contour, we now introduce a probe scalar field. We consider the contours in two and three dimensions discussed earlier. We will argue that, as in the double cone contour, the asymptotic boundary conditions restrict the spectrum to the quasinormal modes of the black hole (and their mirror images in the complex plane, which we will argue are not included in the actual physical spectrum). This is a non-trivial result, particularly in the three-dimensional case, where the modes mix as one goes through the cosmology portion of the contour. This provides evidence that the one-loop determinant for these contours will behave in the same way as in \cite{Chen:2023hra}.

\subsection{Scalar fields in two dimensions}
We consider scalar fields on the contours in the complex plane considered in section \ref{s:ks2d}. We will first consider the double-cone geometry in two dimensions, discussed in \cite{Chen:2023hra}, and then extend the analysis to the contour that passes through the cosmology.

\subsubsection{Scalar fields over the double cone}
In the double cone, following \cite{Chen:2023hra}, we expand the metric (\ref{eq:dcme}) about $u = 0$. In this limit the metric behaves like a $(1+1)$-dimensional Rindler spacetime,
\begin{equation}\label{eq:rind2d}
    \diff s^2 \approx - \rho^2 \, \diff t^2 + \diff u^2.
\end{equation}
Let us consider a minimally coupled massive scalar field over the double cone contour. Our goal is to solve the equations of motion subject to appropriate boundary conditions. We take
\begin{equation}
\phi(t, u) = \ep^{\ic \omega t} g(u)
\end{equation}
so the equations of motion become
\begin{equation}
    \frac{1}{u} \partial_u \qty(u \, \partial_u g)  + \qty(\frac{\omega^2}{r^2} - m^2) g = 0,
\end{equation}
where $m$ is the mass of the scalar field. The approximation (\ref{eq:rind2d}) is only valid near $u = 0$; the solution in this region is
\begin{equation}\label{eq:maldacenawfs}
    g(u) = A_\omega u^{- \ic \omega} + B_\omega u^{\ic \omega}.
\end{equation}

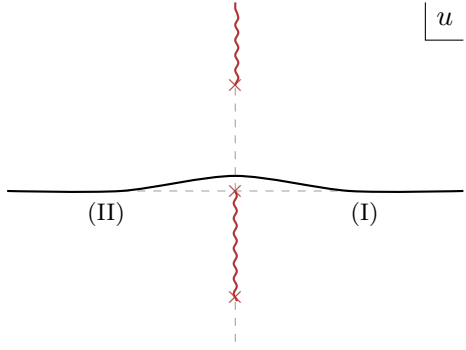
\begin{figure}
    \centering
    \begin{tikzpicture}[scale=1]
        \draw[dashed,mgray] (-1.5,0)--(1.5,0) (0,-2)--(0,-1.4) (0,0)--(0,1.4);
        \draw (2.5,2.5)--(2.5,2)--(3,2) (3,2.5)node[anchor=north east]{$u$};
        \draw[thick] (-1.7,0)node[anchor=north]{\footnotesize (II)} (1.7,0)node[anchor=north]{\footnotesize (I)};
        \draw[thick] plot [smooth] coordinates{(-3,0) (-1.5,0) (0,.2) (1.5,0) (3,0)};
        \draw[decorate,decoration={snake,amplitude=.2mm,segment length=2mm},thick,color=mred] (0,0)node{\footnotesize $\times$}--(0,-1.4)node{\footnotesize $\times$} (0,1.4)node{\footnotesize $\times$}--(0,2.5);
    \end{tikzpicture}
    \caption{Probe scalar over the SSS contour \cite{Saad:2018bqo} in the $u$-plane for the AdS$_2$ black hole.}\label{fig:sss}
\end{figure}

We evolve our solutions from $u>0$ (region I) to $u<0$, (region (II)), along the contour described in figure \ref{fig:sss}, which is the double-cone contour of \cite{Saad:2018bqo}. We denote the radial coordinate in region (I) as $u = \rho_+$, and in region (II) as $u = \ep^{\ic \pi} \rho_-$.

We impose standard asymptotic boundary conditions as $\rho_\pm \to \infty$, keeping only the modes that fall at infinity. Since $\phi_\omega$ is an analytic function away from the branch cut at $u = 0$, we can match the solutions in region (I) and region (II) up to an overall factor that is fixed by the normalization of the mode functions at the two boundaries. Going around the origin, we find
\begin{equation}
   g(\rho_+) = g(\ep^{\ic \pi} \rho_-) = A_\omega \ep^{\pi \omega} \rho_-^{- \ic \omega} + B_\omega \ep^{-\pi \omega} \rho_-^{\ic \omega}.
\end{equation}
To satisfy the boundary conditions at both ends we need $\phi_{\omega}(\rho_-)$ to have the same functional form as $\phi_\omega(\rho_+)$, that is
\begin{equation}
    g(\rho_-) = \Lambda(\omega) g(\rho_+).
\end{equation}
This can be satisfied in three ways:
\begin{enumerate}
    \item $A_\omega = 0$, which corresponds to the quasinormal modes,
    \item $B_\omega = 0$, which corresponds to the anti-quasinormal modes,
    \item $\omega = \ic n$ with $n \in \bbZ$. This does not satisfy the asymptotic boundary conditions unless it also satisfies either of the two conditions above.
\end{enumerate}
Following this logic, \cite{Chen:2023hra} found that these conditions imply that the spectrum of the scalar field on the double cone obtained from this analysis of the wave equation consists of the quasinormal modes and their mirror images in the upper half-plane. We expect only the former to appear in the physical spectrum, as the general Kontsevich-Segal argument leads us to expect that the one-loop determinant will be well-behaved, which it will not be if there are poles in the upper half-$\omega$-plane.\footnote{On the other hand, if we had chosen the contour to go above the singularity i.e., $-\ic \epsilon$ prescription, we would have expected the anti-quasinormal modes to contribute to the one-loop determinant. That is because we are computing $\Tr(\ep^{\ic K T})$ which is well behaved in the limit $T \to \infty$.} The authors of \cite{Chen:2023hra} observed that this double cone contour corresponds to a modified boost operator $\tilde K$ which is part of a one-parameter family of operators obtained by conjugation with the generator $P$, which all have the same spectrum. This family includes the Hamiltonian for the Euclidean AdS$_2$ solution at $u = \ic \frac{\pi}{2}$, showing that the spectrum includes only quasinormal modes.

\subsubsection{Scalar fields over the new contour in two dimensions}
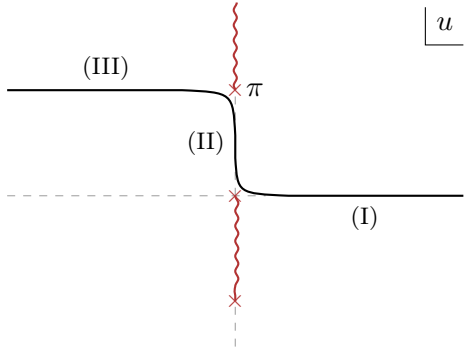
\begin{figure}
    \centering
    \begin{tikzpicture}[scale=1]
        \draw[dashed,mgray] (-3,0)--(.4,0) (0,-2)--(0,-1.4) (0,0)--(0,.4) (0,1)--(0,1.4);
        \draw (2.5,2.5)--(2.5,2)--(3,2) (3,2.5)node[anchor=north east]{$u$};
        \draw[thick] (1.7,0)node[anchor=north]{\footnotesize (I)};
        \draw[thick] (0,.7)node[anchor=east]{\footnotesize (II)} (0,1.4)node[anchor=west]{$\pi$};
        \draw[thick] (-1.7,1.4)node[anchor=south]{\footnotesize (III)};
        \draw[thick] (3,0)--(.7,0)..controls(.03,.03)..(0,.5)--(0,.8)..controls(-.03,1.37)..(-.7,1.4)--(-3,1.4);
        \draw[decorate,decoration={snake,amplitude=.2mm,segment length=2mm},thick,color=mred] (0,1.4)node{\footnotesize $\times$}--(0,2.5) (0,0)node{\footnotesize $\times$}--(0,-1.4)node{\footnotesize $\times$};
    \end{tikzpicture}
    \caption{The cosmological contour in the $u$-plane contributing to the spectral form factor in AdS$_2$. The red lines denote the branch cuts.}\label{fig:2d-ucontour}
\end{figure}

Let us move on to our cosmological contour in two dimensions. The metric is
\begin{equation}
    \diff s^2 = - \sinh^2 u \, \diff t^2 + \diff u^2.
\end{equation}
We now look at the complex $u$ contour in figure \ref{fig:2d-ucontour} that has the following three regions:
\begin{enumerate}
    \item Region (I): $u = \rho_+$, $\rho_+ \in [0, \infty)$, with the line element $\diff s^2 = - \sinh^2 \rho_+ \, \diff t^2 + \diff \rho_+^2$,
    \item Region (II): $u = \ep^{\ic \frac{\pi}{2}} \xi$, $\xi \in [0,\pi]$, with the line element $\diff s^2 = \sin^2 \xi \, \diff t^2 - \diff \xi^2$,
    \item Region (III): $u = \ep^{\ic \pi} \rho_- + \ic \pi$, $\rho_- \in [0, \infty)$, with the line element $\diff s^2 = - \sinh^2 \rho_- \, \diff t^2 + \diff \rho_-^2$.
\end{enumerate}
The metrics in regions (I) and (III) are the same, but for clarity, we use different names for the regions that represent different parts of the contour. Here we have taken regions (II) and (III) with a positive imaginary part of $u$. The analysis with a negative imaginary part for regions (II) and (III) proceeds similarly.

The coordinates in different regions are related to each other as
\begin{equation}
    \rho_+ = \ep^{\ic \frac{\pi}{2}} \xi = \ep^{\ic \pi} \rho_- +\ic \pi.
\end{equation}
With the ansatz $\varphi(t, u) = \ep^{\ic \omega t} g(u)$ the massive scalar equation of motion is
\begin{equation}
    \frac{1}{\sinh u} \partial_u \qty(\sinh u \, \partial_u g) + \qty(\frac{\omega^2}{\sinh^2 u} - m^2) g = 0,
\end{equation}
with solution
\begin{equation}\label{eq:2dcontourwfs}
\begin{split}
    g(u) &= B_\omega \, (\tanh u)^{\ic \omega} (\cosh u)^{\Delta - 1} {}_2F_1\qty[a_+, \frac{1}{2} + a_+, 1 + \ic \omega; \tanh^2 u] \\
    &\quad+ A_\omega \, (\tanh u)^{-\ic \omega} (\cosh u)^{\Delta - 1} {}_2F_1\qty[a_-, \frac{1}{2} + a_-, 1 - \ic \omega; \tanh^2 u],
\end{split}
\end{equation}
where $\Delta = \frac{1}{2} \qty(1 + \sqrt{1 + 4m^2})$ is the conformal dimension and
\begin{equation}\label{eq:hype}
    a_\pm = \frac{1}{2} \qty(1 - \Delta) \pm \frac{\ic}{2} \omega.
\end{equation}
Near $u \to 0$ the solution (\ref{eq:2dcontourwfs}) can be expanded to give the limiting Rindler expansion encountered in (\ref{eq:maldacenawfs}), i.e.,
\begin{equation}
    \lim_{u\to 0} g(u) = A_\omega u^{-\ic \omega} + B_\omega u^{\ic \omega}.
\end{equation}

Now we do the analog of the double-cone matching conditions for our contour in the complex $u$-plane. Our prescription is that we demand a single-valued mode solution while crossing the branch points by matching across distinct contours up to overall normalizations. We apply the same standard boundary condition as before in regions (I) and (III).

We go over the contour shown in figure \ref{fig:2d-ucontour} in the following order: from region (I) to region (II), we rotate in the clockwise direction with $\rho_+ = \ep^{\ic \frac{\pi}{2}} \xi$, and after evolving along the imaginary axis, at $u = \ic \pi$ we rotate in the anticlockwise direction with $\xi = \ep^{\ic \frac{\pi}{2}} \rho_- - \pi$. The first step corresponds to a rotation slightly above $0$ in the complex $u$-plane, followed by a translation along the imaginary axis, and finally another rotation slightly below $u =\ic \pi$. This leads to the solution
\begin{equation}
    \begin{split}
    g\qty(\rho_+) &= \ep^{\ic \pi \Delta} \ep^{-\pi \omega} B_\omega \, (\tanh \rho_-)^{\ic \omega} (\cosh \rho_-)^{\Delta - 1} {}_2F_1\qty[a_+, \frac{1}{2} + a_+, 1 + \ic \omega; \tanh^2 \rho_-] \\
    &\quad+ \ep^{\ic \pi \Delta} \ep^{\pi \omega} A_\omega \, (\tanh \rho_-)^{-\ic \omega} (\cosh \rho_-)^{\Delta - 1} {}_2F_1\qty[a_-, \frac{1}{2} + a_-, 1 - \ic \omega; \tanh^2 \rho_-],
    \end{split}
\end{equation}
Despite the presence of a cosmology segment, this is (apart from an overall phase shift $\ep^{\ic\pi \Delta}$) the same result as the one obtained for the double cone: the coefficients of the two solutions pick up factors of $\ep^{\pm \pi \omega}$ as we evolve from region (I) to region (III).

Since the asymptotics in the two cone regions are the same, and we apply the same boundary condition as in the double cone case, the spectrum of fluctuations around our contour is the same as before: that is, we can satisfy the boundary conditions at both ends only if $A_\omega=0$ or $B_\omega=0$. This gives a spectrum that includes both the quasinormal modes and the anti-quasinormal modes.

As in the double-cone case, we expect that only quasinormal modes are present in the physical spectrum, as the one-loop determinant should be well-behaved by the general Kontsevich-Segal argument. It would be interesting to give a more direct argument for this; the argument based on conjugation from \cite{Chen:2023hra} does not seem to extend straightforwardly to our contour.

\subsection{Scalar fields in three dimensions}
In the three-dimensional case, the geometry is
\begin{equation}
    \diff s^2 = - \rh^2 \sinh^2 u \, \diff t^2 + \rh^2 \cosh^2 u \, \diff \phi^2 + \diff u^2.
\end{equation}
Letting $\varphi = \ep^{\ic \omega \rh t + \ic k \rh \phi} g(u)$, the massive scalar equation of motion is
\begin{equation}
    \frac{1}{\sinh u \cosh u} \partial_u \qty(\sinh u \cosh u \, \partial_u g) + \qty(\frac{\omega^2}{ \sinh^2 u} - \frac{k^2}{\cosh^2 u} - m^2) g = 0.
\end{equation}
The solutions are easy to obtain (see, e.g., \cite{Birmingham:2001pj}). The general solution is
\begin{equation}\label{eq:3d-scalar-sol}
\begin{split}
    g(u) &= B_{\omega,k} \, (\tanh u)^{\ic \omega} (\cosh u)^{\Delta-2}\, {}_2F_1[a_{++}, a_{+-}, 1+\ic \omega; \tanh^2 u] \\
    &\quad+ A_{\omega,k} \, (\tanh u)^{-\ic \omega} (\cosh u)^{\Delta-2}\, {}_2F_1[a_{-+}, a_{--}, 1-\ic \omega; \tanh^2 u],
\end{split}
\end{equation}
where
\begin{equation}
    a_{\epsilon \eta} = \frac{1}{2} (2 - \Delta) + \frac{\ic}{2}(\omega \epsilon + k \eta ), \quad  \epsilon, \eta = \pm 1,
\end{equation}
with $\Delta = 1 + \sqrt{1 + m^2}$.

For the cosmological part of the contour, $u = \ic \xi$, we have
\begin{equation}
\begin{split}
    g(\xi) &= \beta_{\omega,k} \, (\tan \xi)^{\ic \omega} (\cos \xi)^{\Delta-2}\, {}_2F_1\qty[a_{++}, a_{+-}, 1+\ic \omega; -\tan^2 \xi] \\
    &\quad+ \alpha_{\omega,k} \, (\tan \xi)^{-\ic \omega} (\cos \xi)^{\Delta-2}\, {}_2F_1\qty[a_{-+}, a_{--}, 1-\ic \omega; -\tan^2 \xi].
\end{split}
\end{equation}
We relate the solutions near $\xi=0$ to the solutions near $\xi = \frac{\pi}{2}$ using the transformation formula for the hypergeometric functions,
\begin{equation}
\begin{split}
    {}_2F_1(a,b,c;z) &= \frac{\Gamma(c) \Gamma(b-a)} {\Gamma(b) \Gamma(c-a)} (-z)^{-a} {}_2F_1\qty[a, a-c+1, a-b+1; \frac{1}{z}] \\
    &\quad+ \frac{\Gamma(c) \Gamma(a-b)}{\Gamma(a)\Gamma(c-b)} (-z)^{-b}{}_2F_1\qty[b,b-c+1,b-a+1; \frac{1}{z}].
\end{split}
\end{equation}
Applying the formula in our case, using $\xi = \frac{\pi}{2} - \conj{\xi}$, we get
\begin{equation}
\begin{split}
    g\qty(\conj{\xi}) &= \delta_{\omega,k} \, \qty(\tan \conj{\xi})^{\ic k} \qty(\cos \conj{\xi})^{\Delta-2} {}_2F_1\qty[a_{++}, a_{-+}, 1+ik; -\tan^2 \conj{\xi}] \\
    &\quad+ \gamma_{\omega,k} \, \qty(\tan \conj{\xi})^{-\ic k} \qty(\cos \conj{\xi})^{\Delta-2} {}_2F_1\qty[a_{+-}, a_{--}, 1-i k; -\tan^2 \conj{\xi}],
\end{split}
\end{equation}
where
\begin{equation}\label{eq:crel1}
\gamma_{\omega,k} = \qty(\frac{\Gamma(1-\ic \omega)\Gamma(\ic k)}{\Gamma(a_{-+}) \Gamma(1-a_{+-}) } \alpha_{\omega,k} +  \frac{\Gamma(1+\ic \omega)\Gamma(\ic k)}{\Gamma(a_{++}) \Gamma(1-a_{--})}  \beta_{\omega,k}),
\end{equation}
and
\begin{equation}\label{eq:crel2}
    \delta_{\omega,k} = \qty(\frac{\Gamma(1-\ic \omega)\Gamma(-\ic k)}{\Gamma(a_{--}) \Gamma(1-a_{++}) }\alpha_{\omega,k} +  \frac{\Gamma(1+\ic \omega)\Gamma(-\ic k)}{\Gamma(a_{+-}) \Gamma(1-a_{-+})} \beta_{\omega,k}).
\end{equation}
This shows a non-trivial mixing between the solutions across the cosmology in three dimensions, in contrast to the two-dimensional case.

\begin{figure}
    \centering
    \begin{tikzpicture}[scale=1]
        \draw[dashed,mgray] (-3,0)--(.4,0) (0,-2)--(0,-1.4) (0,0)--(0,.4) (0,1)--(0,1.4);
        \draw (2.5,2.5)--(2.5,2)--(3,2) (3,2.5)node[anchor=north east]{$u$};
        \draw[thick] (1.7,0)node[anchor=north]{\footnotesize (I)};
        \draw[thick] (0,.7)node[anchor=east]{\footnotesize (II)} (0,1.4)node[anchor=west]{$\frac{\pi}{2}$};
        \draw[thick] (-1.7,1.4)node[anchor=south]{\footnotesize (III)};
        \draw[thick] (3,0)--(.7,0)..controls(.03,.03)..(0,.5)--(0,.8)..controls(-.03,1.37)..(-.7,1.4)--(-3,1.4);
        \draw[decorate,decoration={snake,amplitude=.2mm,segment length=2mm},thick,color=mred] (0,1.4)node{\footnotesize $\times$}--(0,2.5) (0,0)node{\footnotesize $\times$}--(0,-1.4)node{\footnotesize $\times$};
    \end{tikzpicture}
    \caption{The contour in the $u$-plane for BTZ black hole.}\label{fig:uplanecontour}
\end{figure}

On the contour at $u = \rho_+$, with $\rho_+ \in (0, \infty)$, as we take the limit $\rho_+ \to \infty$, we have two modes behaving as $\ep^{-\Delta \rho_+}$, $\ep^{(\Delta-2)\rho_+}$. The boundary condition sets the latter mode to zero. Using the transformation formula
\begin{equation}
\begin{split}
    {}_2F_1(a,b,c;z) &= \frac{\Gamma(c) \Gamma(c-a-b)}{\Gamma(c-a) \Gamma(c-b)} {}_2F_1[a,b,a+b+1-c;1-z] \\
    &\quad+ \frac{\Gamma(c) \Gamma(a+b-c)}{\Gamma(a) \Gamma(b)} (1-z)^{c-a-b} {}_2F_1[c-a,c-b,1+c-a-b;1-z],
\end{split}
\end{equation}
we obtain
\begin{equation}
    \frac{\Gamma(1- \ic \omega)}{\Gamma(1-a_{++})\Gamma(1 - a_{+-})} A_{\omega,k}  + \frac{\Gamma(1 + \ic \omega)}{\Gamma(1-a_{-+})\Gamma(1-a_{--})} B_{\omega,k} = 0.
\end{equation}
Under the contour rotation near $u=0$, we get $A_{\omega,k} = \ep^{-\frac{\pi}{2} \omega} \alpha_{\omega,k}$, $B_{\omega,k} = \ep^{\frac{\pi}{2} \omega} \beta_{\omega,k}$. Consequently, we have
\begin{equation}\label{eq:3d-ab}
\frac{\alpha_{\omega,k}}{\beta_{\omega,k}} = \ep^{\pi \omega} \frac{A_{\omega,k}}{B_{\omega,k}} = - \ep^{\pi \omega} \frac{\Gamma(1+\ic \omega)\Gamma(1-a_{++})\Gamma(1 -a_{+-}) } {\Gamma(1-\ic \omega)\Gamma(1-a_{-+})\Gamma(1 -a_{--}) }
\end{equation}

Similarly, on the contour at $u = \ic \frac{\pi}{2} - \rho_-$, the solution is given by
\begin{equation}
\begin{split}
    g &= D_{\omega,k} \, (\tanh \rho_-)^{\ic k} (\cosh \rho_-)^{\Delta-2} {}_2F_1\qty[a_{++}, a_{-+}, 1+\ic k; \tanh^2 \rho_-] \\
    &\quad+ C_{\omega,k} \, (\tanh \rho_-)^{-\ic k} (\cosh \rho_-)^{\Delta-2} {}_2F_1\qty[a_{+-}, a_{--}, 1-\ic k; \tanh^2 \rho_-].
\end{split}
\end{equation}
Imposing the asymptotic boundary condition on this contour gives
\begin{equation}
\frac{\Gamma(1-\ic k)}{\Gamma(1-a_{++})\Gamma(1 - a_{-+})} C_{\omega,k} + \frac{\Gamma(1+\ic k)}{\Gamma(1-a_{+-})\Gamma(1 -a_{--})} D_{\omega,k} = 0.
\end{equation}
The contour rotation near $\rho_-=0$ is $\rho_- = \ic \conj{\xi}$, which gives $C_{\omega,k} = \ep^{-\frac{\pi}{2} k} \gamma_{\omega,k}$, $D_{\omega,k} = \ep^{\frac{\pi}{2} k} \delta_{\omega,k}$, so we get
\begin{equation}\label{eq:bc2}
    \frac{\gamma_{\omega,k}}{\delta_{\omega,k}} = \ep^{\pi k} \frac{C_{\omega,k}}{D_{\omega,k}} = - \ep^{\pi k} \frac{\Gamma(1+\ic k)\Gamma(1-a_{++})\Gamma(1 - a_{-+} )} {\Gamma(1-\ic k)\Gamma(1-a_{+-})\Gamma(1 -a_{--}) },
\end{equation}
while (\ref{eq:crel1}, \ref{eq:crel2}, \ref{eq:3d-ab}) give
\begin{equation}\label{eq:bgrel}
    \frac{\gamma_{\omega,k}}{\delta_{\omega,k}} = - \ep^{-\pi k}\frac{\Gamma(1+\ic k)\Gamma(1-a_{++})\Gamma(1 - a_{-+} )} {\Gamma(1-\ic k)\Gamma(1-a_{+-})\Gamma(1 -a_{--}) }.
\end{equation}
Thus, if $\ep^{2\pi k} \neq 1$, and the above analysis holds, $\omega,k$ are not in the spectrum.

The cases that could survive are where $\ic \omega$, $\ic k$, or one of the $a_{\pm\pm}$ is an integer, where some of the Gamma functions above have poles, and the general analysis breaks down. In fact, only the cases where one of the $a_{\pm\pm}$ is a positive integer survive. 

In the cases where $\ic \omega$ or $\ic k$ is an integer, we get only one linearly independent solution in terms of a hypergeometric function. The other solution will, in general, involve $\log u$ near $u = 0$ or $u = \ic \frac{\pi}{2}$, and as in \cite{Chen:2023hra} it is not possible to satisfy the boundary conditions at both ends for these solutions with logarithmic behavior. If $a_{\pm\pm}$ is zero or a negative integer, there is no simplification in the asymptotic boundary conditions; one of the relations (\ref{eq:crel1}) and (\ref{eq:crel2}) simplifies in these cases, but the inconsistency between equations (\ref{eq:bc2}) and (\ref{eq:bgrel}) persists.

This leaves the cases where one of $a_{\pm\pm}$ is a positive integer. If, for example, $a_{++}$ is a positive integer, the boundary conditions simplify to $B_{\omega,k}=0$ and $D_{\omega,k}=0$, and these are consistent conditions as $\delta_{\omega,k} \propto \beta_{\omega,k}$. Similarly for the other cases. The mismatch between equations (\ref{eq:bc2}) and (\ref{eq:bgrel}) is resolved as these are both equal to zero or infinity, so the mismatch in the multiplying factor is irrelevant.

We conclude that the allowed values of $\omega$ are
\begin{equation}
\omega = \pm k \pm \ic (2m + \Delta), \quad m \in \bbZ_{\geq 0},
\end{equation}
where the two sign choices are independent. The positive sign in the second term corresponds to the quasinormal modes obtained in \cite{Birmingham:2001pj}. Thus, we once again find that the spectrum of fluctuations around this contour is given by the quasinormal modes and their mirror images in the upper half-plane. This is particularly striking in this case, where the form of the solutions along the two cone portions of the contour is different, and hence the boundary conditions differ; this is compensated by the non-trivial transformation across the cosmology. As in the previous cases, we expect that the actual physical spectrum will be just the quasinormal modes. It would be interesting to give a more direct argument for this, but we leave that for future work.

% -------------------------
% discussion
% -------------------------
\section{Discussion and speculations}\label{s:discussion}
The torus cosmology 
\begin{equation}
\diff s^2 = -\diff \xi^2 + \rh^2 \sin^2\xi\, \diff t^2 + \rh^2 \cos^2 \xi\, \diff \phi^2
\end{equation}
with $\phi\sim\phi + 2\pi$ and $t\sim t+T$ is a natural solution of Einstein gravity with a negative cosmological constant in three dimensions. In this paper we have argued that the big bang/crunch singularities at $\xi=0$ and $\xi = \frac{\pi}{2}$ can be resolved by taking a small detour in the complex plane, connecting the cosmology to external cone regions. Although the region where the metric is complex can be infinitesimally small, it leads to a surprising result: the amplitude for this cosmology is, in a holographic theory, computed by the spectral observable 
\begin{equation}\label{eq:3d-ramplitude}
    \overbar{Y_{\tilde{E},\Delta \tilde{E}}(\tilde{T}) Y_{E, \Delta E}(-T)} \simeq T
\end{equation}
where $E=\frac{\rh^2}{8G}$, $\tilde{E} = \frac{T^2}{4\pi^2} E$ and $\tilde{T} = \frac{4\pi^2}{T}$. We are using units where $\ell_\AdS = 1$. The left-hand side of this equation is an observable in the boundary conformal field theory; it is the two-point function of the density of states, which describes correlations between black hole states with different energy. The right-hand side is a prediction based on the computation of the action at classical and one-loop order. It is important to note that the prediction here is for the \textit{connected} part of the correlation function.  Of course, there will be disconnected contributions as well, which---as with the double cone---are simply related to the averaged density of states.

Several comments are in order.  First, as we vary $T$, it is impossible to keep both $E$ and $\tilde{E}$ fixed. So, this is not a spectral observable that describes correlations between black hole microstates with fixed energy. One option is to take $E$ fixed and $T$ large, in which case it describes correlations between states with finite energy and those with very large energy.  Another option is to keep $\tilde{E}$ fixed as $T$ becomes large, in which case it describes correlations between black hole states and those with $E\to 0$.  This option is interesting because the limit $E \to 0$ describes small black holes, i.e., those just above the threshold for black hole formation in AdS$_3$.

This way of interpreting the formula (\ref{eq:3d-ramplitude}) may provide important information about the dual CFT. In CFT language, the states on the circle are dual to operators with scaling dimension $\Delta = E + \frac{c}{12}$ (recall that we are measuring $E$ in AdS units). If we take large $T$ with fixed $\tilde{E}$, this means that we are taking $\rh \sim \sqrt{G}/T \to 0$. In the CFT language, this means that we are considering the correlations of black hole states (which have dimension $\Delta \gg \frac{c}{12}$) with those of dimension $\Delta = \frac{c}{12}+\cO\qty(\frac{1}{T^2})$.  It is important to note that, in order to trust our computation, we should require that $\rh \sim \sqrt{G}/T \gg G$, so that the ``light'' black holes under consideration are still large in Planck units. So we must keep $T \ll 1/\sqrt{G}$. This means that our prediction describes correlations between heavy states and those with $\Delta - \frac{c}{12} \gg \cO\qty(\frac{1}{c})$. In the semi-classical (large $c$) limit this includes states which are close to the threshold for black hole formation. If we impose the stronger condition that the light black holes have horizons which are AdS scale, i.e., $\rh = \cO(1)$, we find that the light states under consideration have $\Delta - \frac{c}{12} = \cO\qty(c)$.

This is an important regime from the point of view of the dual CFT because it is one where the structure of the density of states may depend on the detailed nature of the microscopic theory.  In particular, it includes states that lie in the so-called ``enigmatic regime'' $\frac{c}{12}<\Delta<\frac{c}{6}$.  As shown in \cite{Hartman:2014oaa}, simply imposing modular invariance along with a sparse light spectrum does not determine the density of states in this regime at large $c$. 
We see that the structure of states in this regime now has a cosmological interpretation. It is tempting to speculate that equation (\ref{eq:3d-ramplitude}) represents the behavior in a ``typical'' holographic theory, and that in such a theory the big bang/crunch singularities may be regulated by complexifying the geometry. In an atypical theory, this may not be the case.\footnote{Similar statements can be made about the double-cone geometry, which predicts a linear ramp in the spectral form factor expected in a typical (chaotic) conformal field theory.  In special theories, including those that are integrable, the ramp may not be present.}

A second comment is that, so far, we have been frustratingly vague about the exact nature of the averaging implied by the overline in equation (\ref{eq:3d-ramplitude}).  It would be interesting to understand whether there is a natural ensemble of two-dimensional CFTs (e.g., as in \cite{Chandra:2022bqq}) where the statistical properties of the ensemble reproduce this prediction.  In pure gravity, some insight might be gained by considering the Euclidean $\bbT^2 \times \bbR$ wormholes considered in \cite{Cotler:2020ugk}.  Or it may be interesting to consider restricted ensembles of CFTs where the averaging can be carried out explicitly, as in \cite{Maloney:2020nni, Afkhami-Jeddi:2020ezh}; in fact, the two-point function of the density of states in the ensemble of Narain CFTs was computed in \cite{Collier:2021rsn}. This could lead to a cosmological interpretation of the observables of $\rU(1)$ gravity.
Of course, it may also be possible to clarify this by considering string theories in AdS$_3$.  The big bang/crunch singularities that arise in our torus cosmology are very similar to those considered in the string theory constructions of \cite{Liu:2002yd,Elitzur:2002vw}.\footnote{Indeed, the asymptotically AdS$_3$ regions which we obtain by smoothing out the cosmological singularity appear to be closely related to the ``whiskers'' discovered in the coset CFTs of \cite{Elitzur:2002vw}.  We thank J. Maldacena for pointing this out to us.}  This would be interesting to pursue in future work.

Third, we have so far made comments only about the statistical properties of the density of states but not about other CFT data, such as OPE coefficients.  There are natural  generalizations of the torus wormhole of the form
\begin{equation}
    \diff s^2 = -\diff \xi^2 + \sin^2 \xi \, \diff \Sigma^2
\end{equation}
where $\diff \Sigma^2$ is the metric on a constant negative curvature surface.  This metric can be generalized so that the geometry of this surface depends on $\xi$, e.g., as in \cite{Krasnov:2005dm}. We propose that one can similarly resolve the cosmological singularities of this space-time by complexifying the geometry and that the result will take the form of a statistical prediction for the properties of OPE coefficients in the ``enigmatic regime.'' This would be interesting to pursue but is more technically challenging than the present computation.

Fourth, we note that the prediction (\ref{eq:3d-ramplitude}) is rather special to two-dimensional CFTs.  As we discuss in appendix \ref{s:kshd}, there does not seem to be a straightforward higher-dimensional analog.

Finally, using the modular invariance of two-dimensional CFTs one can give different interpretations to our spectral observable. To apply a modular transformation, one needs to consider a more conventional spectral form factor involving the canonical partition function $Z(\beta+\ic T)$. There are two subtleties when doing this: first, it was observed in \cite{Saad:2018bqo} that the double cone is not really a saddle point for the canonical spectral form factor, due to a pressure towards lower energies. Second, for states that are close to the BTZ threshold, it is not clear whether the usual relation between energy and temperature will be valid. However, there is at least one case where it seems legitimate.  Let us consider equation (\ref{eq:3d-ramplitude}) at large $T$ and fixed $E$ so that $\tilde{E}$ is becoming very large.  So we expect that we can freely interpret $Y_{\tilde{E},\Delta \tilde{E}}(\tilde{T})$ in terms of a canonical ensemble partition function.  We can then perform a modular transformation to relate this to low-energy states.  In this case our observable (\ref{eq:3d-ramplitude}) is interpreted as a connected correlation function between black holes and the light spectrum of the theory (i.e., those states with $\Delta \ll \frac{c}{12}$). We can then ask what sort of light states could contribute to this connected correlator. The light spectrum includes empty AdS (i.e., the identity operator) and its Virasoro descendants.  This spectrum is fixed entirely by symmetry and therefore should not contribute to a connected correlator.  So our observable is interpreted as describing correlations between the spectrum of black hole microstates and light, non-vacuum primary states in the theory.

This may have important implications for pure gravity.  In particular, a ``pure'' theory of quantum gravity (in the sense of \cite{Witten:2007kt}) is one that possesses no primary states other than the vacuum in the window $-\frac{c}{12} < E < 0$.  A theory with no states in this window cannot have non-trivial spectral correlations with black hole microstates. {{Simply the existence of a nonzero connected correlation function would seem to rule out the existence of such pure theories of quantum gravity!}} In order to make this argument precise, however, one needs a clearer understanding of the precise circumstances, and for what CFTs, equation (\ref{eq:3d-ramplitude}) will hold.  A more conservative conclusion would be that, if pure gravity exists, it is highly non-generic.  Perhaps for some reason, multi-boundary saddle points do not contribute to the path integral. This would be expected in integrable theories, which do not have the usual ramp described by the double cone.\footnote{Indeed, the candidate pure theory of gravity related to the Monster CFT discussed in \cite{Witten:2007kt} is integrable, in the sense that every state in the theory is a descendant of the vacuum under an appropriate chiral algebra.}  In any case, our argument implies that, if a pure theory of quantum gravity exists, its saddle points likely do not include the complexified torus cosmology we have described above.

% -------------------------
% acknowledgments
% -------------------------
\section*{Acknowledgements}
We thank X. Dong, T. Hartman, J. Maldacena, S. Shenker, and especially C. Johnson for useful conversations, and M. Sasieta and O. Janssen for comments on previous versions of the paper. The work of J.C. is supported by Simons Collaboration on the Nonperturbative Bootstrap. S.F.R. is supported in part by STFC through grant ST/T000708/1. A.M. and K.N. are supported in part by the Natural Sciences and Engineering Research Council of Canada (NSERC), funding reference number SAPIN/00047.

% -------------------------
% appendix
% -------------------------
\appendix
\section{Contours in higher dimensions}\label{s:kshd}
In section \ref{s:contours}, we considered contours in the complex $u$-plane in two and three dimensions. It is natural to ask what the analogs are in higher dimensions. It turns out that the previous discussion is quite special to low dimensions. One significant reason is that in the BTZ case, region (III) in figure \ref{fig:3d-ucontour} involves flipping a minus sign for the metric components $g_{tt}$ and $g_{\phi \phi}$ as compared to region (I). For the AdS$_{d+1}$ black hole with $d>2$, we have $d>1$ transverse directions, and consequently, the contour in region (III) does not satisfy the KS criterion (\ref{eq:ksc}).

Nevertheless, it might be interesting to understand what kind of contours we can have in higher dimensions. Consider for simplicity AdS$_5$-Schwarzschild with the following metric:
\begin{equation}
    \diff s^2 = -f(r) \, \diff t^2 + \frac{\diff r^2}{f(r)} + r^2 \diff \Omega_3^2, \qquad f(r) = \frac{r^2}{\ell^2} - \frac{\rh^4}{\ell^2 r^2}.
\end{equation}
The analog of the $u$-coordinate used previously is an isotropic coordinate, covering the whole Einstein-Rosen bridge. With $r = \rh \sqrt{\cosh u}$, we have
\begin{equation}
    \diff s^2 = - \frac{\rh^2 \sinh^2 u}{\ell^2 \cosh u} \, \diff t^2 + \frac{\ell^2}{4} \, \diff u^2 + \rh^2 \cosh u \, \diff \Omega_3^2.
\end{equation}
The Einstein-Rosen bridge is given by real $u \in (-\infty,\infty)$. The region with $u = \ic \xi$ is a cosmology, propagating from a big bang at $\xi = 0$ to a big crunch at $\xi = \frac{\pi}{2}$. The novelty here is that the proper length in the $t$ direction blows up at $\xi = \frac{\pi}{2}$; this is associated with the negative power of $r$ in $f(r)$.

As expected, the horizontal contour at $u = \rho + \ic \frac{\pi}{2}$ is not allowed. The time and sphere directions have a factor of $\ic$ here, so $ \sum_{\mu=1}^5 \abs{\Arg{\lambda_\mu}} = 2\pi$ along this curve. In general, writing $u=x+\ic y$,
\begin{equation}
    \sum_{\mu=1}^5 \abs{\Arg{\lambda_\mu}} = 2 \arctan \abs{\frac{\diff y}{\diff x}} + 4 \arctan (\tan y \tanh \abs{x}) + 2 \arctan( \tan (\frac{\pi}{2} - y)  \tanh \abs{x}).
\end{equation}

\begin{figure}
    \centering
    \input{5d-contours.pgf}
\caption{The limit of all allowed contours for AdS$_5$-Schwarzschild.}\label{fig3}
\end{figure}

There is a region where the RHS is $>\pi$ even for horizontal contours, where there are no allowed contours. The boundary of this region is at $\tan^2 y = \coth^2 x - 2$, which is plotted in figure \ref{fig3}. We see that there are no contours that satisfy the KS condition which reach large $x$. The real axis, $u=x$, has $ \sum_{\mu=1}^5 \abs{\Arg{\lambda_\mu}} = \pi$, on the margin of satisfying the KS condition, but for $ \coth^2 x < 2$ deforming away from the real axis increases $ \sum_{\mu=1}^5 \abs{\Arg{\lambda_\mu}}$ (near the origin, for $ \coth^2 x >2$, it decreases it). 

This provides a significant obstruction to constructing contours that satisfy the KS condition in higher dimensions, implying also that the double cone contour of \cite{Saad:2018bqo} does not satisfy this condition in higher dimensions.  

% references
\bibliographystyle{JHEP}
\bibliography{refs}
\end{document}